\documentclass[aps,pre,
 reprint,
 twocolumn,amsmath,amssymb,floatfix,
 longbibliography]{revtex4-2}

\usepackage{dcolumn}% Align table columns on decimal point
\usepackage{bm}% bold math
\usepackage[T1]{fontenc}
\usepackage{mathptmx}
\usepackage{etoolbox}

\usepackage[utf8]{inputenc}
\usepackage{amsmath}

\usepackage{graphicx} 
\usepackage[colorlinks=true]{hyperref}

\usepackage{color,soul}
\definecolor{lb}{rgb}{.90,.95,1}
\setulcolor{red}    % set underlining color
\setstcolor{blue}  % set overstriking color
\sethlcolor{yellow}   % set highlighting color
\DeclareUnicodeCharacter{393}{$\Gamma$}

\begin{document}

\title{Simulations and integral--equation theories for dipolar density interacting disks}

\date{\today}

\author{Elena Rufeil--Fiori}
              \email{elena.rufeil@unc.edu.ar} 
\affiliation{Facultad de Matemática, Astronomía, Física y Computación, Universidad Nacional de C\'ordoba, C\'ordoba X5000HUA, Argentina}
\affiliation{Instituto de Física Enrique Gaviola, CONICET - UNC, Córdoba X5000HUA, Argentina}
           
\author{Adolfo J. Banchio}
              \email{ajbanchio@unc.edu.ar}   
\affiliation{Facultad de Matemática, Astronomía, Física y Computación, Universidad Nacional de C\'ordoba, C\'ordoba X5000HUA, Argentina}
\affiliation{Instituto de Física Enrique Gaviola, CONICET - UNC, Córdoba X5000HUA, Argentina}

%
%                 ABSTRACT
%
\begin{abstract}
Integral equation theories (IETs) based on the Ornstein--Zernike (OZ) relation can be used as an analytical tool to predict structural and thermodynamic properties and phase behavior of fluids with low numerical cost. However, there are no studies of the IETs for the dipolar density interaction potential in 2D systems, a relevant inter--domain interaction in lipid monolayers with phase coexistence. 
This repulsive interaction arises due to the excess dipole density of the domains, which are aligned perpendicular to the interface.
This work studies the performance of three closures of the OZ equation for this novel system: Rogers--Young (RY), Modified Hypernetted Chain (MHNC), and Variational Modified Hypernetted Chain (VMHNC). For the last two closures the bridge function of a reference system is required, being the hard disk the most convenient reference system. Given that in 2D there is no analytical expressions for the hard disk correlation functions, two different approximations are proposed:
%Two different approximations for the hard disks reference system are considered for the last two closures, which approximate the bridge function by that of a reference system:
one based on the Percus--Yevick approximation (PY), and the other based on an extension of the hard spheres Verlet--Weis--Henderson--Grundke parameterization (LB). The accuracy of the five approaches is evaluated by comparison of the pair correlation function and the structure factor with Monte Carlo simulation data. 
The results show that RY closure is only satisfactory for low--structured regimes. MHNC and VMHNC closures perform globally well and there are no significant differences between them. However, the reference system in some cases affects their performance; when the pair correlation function serves as the measure, the LB--based closures quantitatively outperform the PY ones. 
From the point of view of its applicability, LB--based closures do not have a solution for all studied interaction strength parameters, and, in general, PY--based closures are numerically preferable.
%The results show that they perform globally quite well and that there are no significant differences between MHNC and VMHNC. The reference system slightly affects the performance. When the pair correlation function serves as the measure, the LB--based closures quantitatively outperform the PY ones. However, LB--based closures do not have a solution for all studied interaction strength parameters, limiting their applicability. On the other hand, PY--based closures are numerically preferable. Finally, the RY closure is satisfactory only for low--structured regimes.

%
\end{abstract} 

\maketitle

%%%MAIN TEXT%%%%

%%%%%%%%%%%%%%%%%%%%%%
\section{Introduction}

In the context of lipid monolayers with phase coexistence, electrostatic inter--domain interactions~\cite{andelman85,andelman86,ursell09} in two-- and quasi--two--dimensional systems play an important role in the lateral organization of the monolayer~\cite{galassi2021}. These inter--domain interactions affect the motion of domains~\cite{rufeil18}, as well as that of other molecules present in the monolayer~\cite{forstner08,ruckerl08}.
Among these interactions, the dipolar density repulsion between domains is particularly relevant because it is always present. It arises from the excess dipolar density, which is perpendicular to the monolayer plane, of the ordered phase of the domains with respect to the continuous phase of the surroundings. Usually, this interaction is simplified by adopting the point dipole approximation~\cite{rufeil16,rufeil18,khrapak2018, hoffmann2006a}. 
This approach, however, is only valid for very dilute systems~\cite{rufeil16}.
%\tm{Sacar: This approach can be convenient since the dipole density interaction lacks a simple analytical expression, and its calculation involves several numerical integrations. Y agregar algo de Sin embargo el dipolo puntual solo vale para sistemas muuuy diluidos, por ejemplo en nuestro trabajo 2016 se ve que difieren mucho.}

Integral equation theories (IETs) of the fluid state represent an alternative, analytical tool to predict structural and thermodynamic properties as well as the phase behavior of fluids~\cite{hansen, caccamo1996, pellicane2020}. Besides, they provide a numerically low-cost approach in comparison with numerical simulations. 

The most used IETs are those based on the Ornstein-Zernike (OZ) relation  associated with an approximate closure relation (OZ--IETs). Previous works in the literature have explored the performance of a few different closures to the OZ equation for the point--dipole interaction in two--dimensional (2D) systems~\cite{hoffmann2006a, hoffmann2006b, vanteeffelen2008}. 
However, as far as we know, there are no OZ--IETs studies on the dipolar density interaction. Because membrane systems are ubiquitous in biological complexes, finding closures to the OZ equation for the 2D dipolar density interaction is essential to develop theoretical models that efficiently predict their thermodynamic properties.
 
Adequate closures to the OZ equation can help us understand how long--range dipolar interactions affect the structure of the system. 
They can also be used to conceive new computational methods and algorithms to simulate these models and predict, for instance, the dipolar density strength of a monolayer~\cite{rufeil16}. 
In mode--coupling schemes~\cite{goetze2008,Szamel1991, naegele1997a,naegele1997b}, used for the study of the dynamics and glass transition, and density functional theory~\cite{vanteeffelen2006,vanteeffelen2008}, chosen for studying inhomogeneous fluids, two--particle correlation function of the systems under consideration are needed. For this reason, these theories benefit from an IET capable of generating accurate pair correlation functions. 
Finally, an adequate closure can also be used in an inverse protocol \cite{heinen2018} to study new inter-domain pair potentials that could be acting in lipid monolayers.  

In this work, we study three main closures of the OZ equation: Rogers--Young (RY), Modified Hypernetted Chain (MHNC), and Variational Modified Hypernetted Chain (VMHNC). 
MHNC and VMHNC closures depend on a reference system, being Hard Spheres (HS) the most used in 3D systems. In 2D there is no analytical expression for the correlation functions of Hard Disks (HD) system. Hence, for each studied closure we consider two different approximations for the HD reference system; one based on the Percus–Yevick approximation (PY) and the other based on HD extension of the HS Verlet-–Weis–-Henderson-–Grundke (VWHG) parameterization~\cite{verlet1972, henderson1975}, suggested by Law and Buzza~\cite{law2009}, that we name LB. 
Thus we are left with five different approaches whose accuracy is evaluated via direct Monte--Carlo simulations. 

Based on the comparison with the simulations, we have found that there is no major difference in performance between MHNC and VMHNC. 
We found that the reference system influences on the accuracy of the results. When measured by the pair correlation function, the LB consistently represents a better choice than PY. However, when the structure factor is used as the measure of accuracy, the reference system that performs better depends on the particular region of the phase diagram.
The closures based on LB do not present a solution for all the studied interaction strength parameter range, which limits the applicability of this reference system. 
Finally, for 
%low interaction strength regime 
low structured systems RY presents the better performance. However, its accuracy worsens with increasing interaction strength.

The paper is organized as follows: In Sec. II, we present the inter--domain interaction, the reduced units, and the structural quantities of interest. 
In Sec. III, we introduce the OZ equation and the closure relations that we considered to study the performance of the OZ--IETs for a 2D--system of hard disks with dipolar density interaction pair--potential.  We also describe the basic aspects of numerical calculations. In Sec. IV, we discuss our results benchmarking our IETs approach with Monte-Carlo simulations. Finally, Sec. V provides our conclusions.

%%%%%%%%%%%%%%%%%%%%%%
\section{Description of the Model}

%%%%%%%%%%%
\subsection{Dipolar density potential}

Dipolar density potential arises naturally in lipid monolayers with its two--phase, liquid--condensed (LC) and liquid--expanded (LE), coexistence region. Here, in general, the LC phase forms domains in the LE phase, which occupies the larger area of the monolayer.
Because of the difference in surface densities, the LC domains possess an excess dipolar density with respect to the surrounding LE phase \cite{mcconnell}, resulting in an inter--domain dipolar repulsive interaction.

\begin{figure}[h]
\centerline{
\includegraphics[width=1.0\columnwidth]{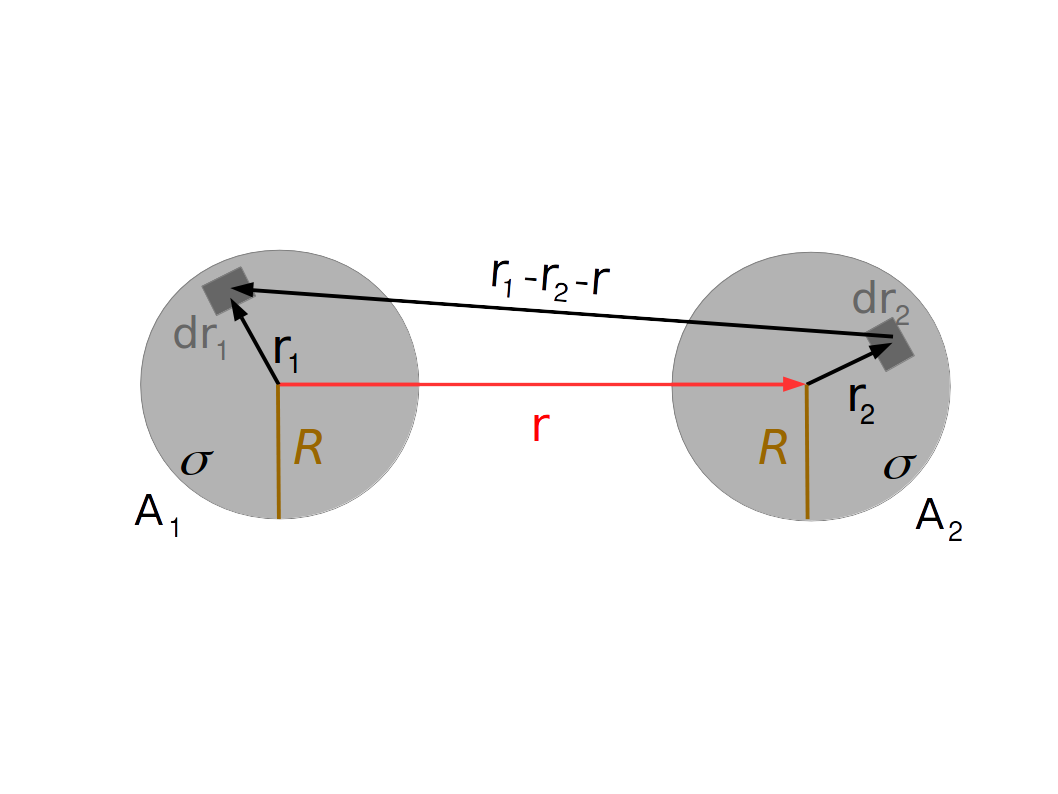} }
\caption{Two domains of equal radii $R$ and excess dipolar density $\sigma$ with center--to--center distance $r$.}
\label{esquema}
\end{figure}

In order to model the mixed monolayer, we consider it as a uniform layer with permittivity $\epsilon_m$ that lies between two different semi--infinite uniform media (air and water) with permittivities $\epsilon_a$ and  $\epsilon_w$, respectively.
This layer is composed of a 2D monodisperse dispersion of circular domains of radii $R$, with condensed area fraction $\phi= N \pi R^2/A$, being $N$ the number of domains and $A$ the monolayer area.  
Each domain possesses an effective dipolar density $\sigma$ perpendicularly oriented to the interface.
Merging mechanism between domains is not considered in this model, and hence, we have added a hard--core potential to prevent it. 
The resulting pair potential between domain 1 and domain 2 can be described by: 
\begin{equation}
\label{Ut}
U_t(r) = U_{hc}(r) + U_{d}(r),
\end{equation}
where $U_{hc}(r)$ is the hard--core repulsive potential, and
$U_d(r)$ is the dipolar density potential described by: 
\begin{equation}
\label{Ud}
 U_d(r)
=f_0\int_{A_2}\int_{A_1}\frac{\mathrm{d}{\bf r}_1\; \mathrm{d}{\bf r}_2 }{|{\bf r}_1-{\bf r}_2-{\bf r}|^3},
\end{equation}
where $A_i$ denotes the area of domain $i$, $d\bf{r}_i$ its area element and $\bf{r}_i$ its position vector respect to the domain center, with $i=1,2$. $\bf{r}$ is the vector from the center of domain 1 to the center of domain 2, as shown in Fig.~\ref{esquema}. Here, we define the interaction strength,
\begin{equation}
\label{f0}
f_0=\frac{\sigma^2}{4\pi\epsilon_0\epsilon^*}, 
\end{equation}
with $\epsilon_0$ the vacuum permittivity and $\epsilon^*$ an effective permittivity~\cite{urbakh93}, $\epsilon^*=\epsilon_m^2(\epsilon_w+\epsilon_a)/(2\epsilon_w\epsilon_a)$. In Eq. (\ref{Ud}) the potential constant was chosen such that the potential tends to zero for infinite separation.

There is no closed analytic expression for $U_d(r)$, and hence it must be calculated numerically. 
However, for the particular case of monodisperse systems (all domain radii equal to $R$), Wurlitzer \textit{et al.}~\cite{fischer02} found the asymptotic behavior of this potential as:
\begin{equation}
\label{Ud_as}
\frac{U_d(r)}{f_0 R} \approx \left\{ \begin{array}{lcr}
             - 4\pi \sqrt{r/R-2} + C  &   & 0 < r/R - 2 \ll 1 \\
             \pi^2 \frac{R^3}{r^3} &   & r/R - 2 \gg 1 \\
             \end{array}
   \right.
\end{equation}
where the contact value, $C$, is obtained from the numerical solution of Eq.~(\ref{Ud}) with $r=2R$, resulting $C\approx5.74216$ (Appendix~\ref{integral_density}). Note that the derivative of the potential diverges when the domains approach contact.

As expected, for large distances it reduces to the interaction of two point--dipoles with dipole moment $\mu=\sigma \pi R^2$:
\begin{equation}
\label{Up}
U_p=\frac{\mu^2}{4\pi\epsilon_0\epsilon^*}\frac{1}{r^3}.
\end{equation}

\begin{figure}[h]
\centerline{
\includegraphics[width=1.0\columnwidth]{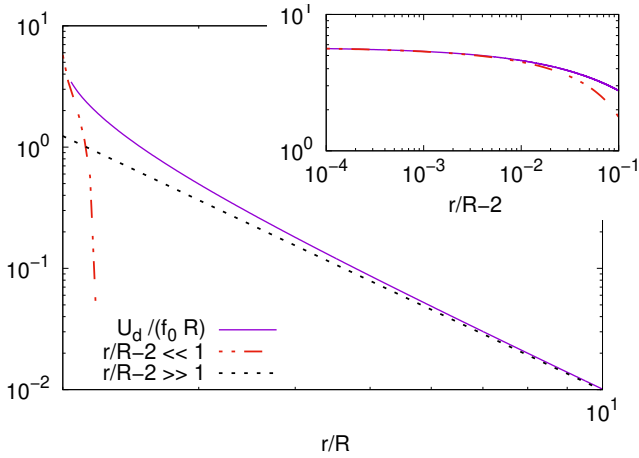}}
\caption{Dipolar density interaction potential, $U_d$, (solid violet line) and the asymptotic expressions for short (dotted--dashed red line) and long distance (dashed black line) given in Eq.~(\ref{Ud_as}). In the inset, the asymptotic expression for short distances is also shown as a function of the border--to--border separation $r/R-2$.}
\label{potential}
\end{figure}

In Figure~\ref{potential} we show the numerical solution of Eq.~(\ref{Ud}) (solid line) and the asymptotic expressions for short (dotted line) and long distance (dashed line) given in Eq.~(\ref{Ud_as}). The short distance asymptotic expression approximates $U_d$ up to $r/R=0.0004$ within an error of $0.1\%$, while the long distance asymptote, the point--dipole approximation, is a good approximation only for $r/R>40$ within the same error.

In certain regions of the monolayer phase diagram, the characteristic length scale of the system is the mean geometrical distance between domains $r_m= \rho^{-1/2}=\sqrt{\pi/\phi}R$, where $\rho=N/A$ is the number density. In these regions, the first neighbor shell is close to the distance $r_m$.

Using $r_m$ as unit length, the dimensionless dipolar potential takes the form:
\begin{equation}
\label{Ud_adim}
 \frac{U_d(\tilde{r})}{k_B T} =  \frac{\Gamma}{\phi^2}\int_{\tilde{A_2}}\int_{\tilde{A_1}}\frac{\; \mathrm{d}\tilde{\bf r}_1 \;\mathrm{d}\tilde{\bf{r}}_2}{|\tilde{\bf r}_1-\tilde{\bf r}_2-\tilde{\bf r}|^3}
\end{equation}
where $\tilde{\bf r}={\bf r}/r_m$,  $\tilde{A_i}=A_i/r_m^2$,  and $\Gamma$ is a dimensionless interaction strength parameter:
\begin{equation}
\label{Gamma}
 \Gamma = \frac{f_0\pi^2R^4}{k_B T\;r_m^3}.
\end{equation}
In these units, the asymptotic point dipole behavior, Eq.~(\ref{Ud_as}), takes the form
\begin{equation}
\label{}
 \frac{U_d(\tilde{r})}{k_B T} \rightarrow \frac{\Gamma}{\tilde{r}^3},\quad\;\; \tilde{r}\gg \frac{R}{r_m}.
\end{equation}
%\frac{R}{r_m}=\sqrt{\frac{\phi}{\pi}}
In dusty plasmas~\cite{bonitz2010} and point dipole monolayers~\cite{hoffmann2006a,vanteeffelen2008} these reduced units are frequently used because in those systems the resulting dimensionless strength parameter (analogous to $\Gamma$) describes the full system.

The advantage of using the mean interparticle distance as the unit of length in the dipolar density interaction is that the three--parameter space $\{f_0, \rho, R\}$ is mapped to a two--parameter space $\{\Gamma, \phi\}$. 
Note that, in this parameter space, the limit $\Gamma \rightarrow 0$ keeping $\phi$ constant presents the subtlety that the limit $f_0 \rightarrow \infty$ is implicit (due to the fact that $\Gamma$ and $\phi$ both depend on $\rho$). The same happens with the limit $\phi \rightarrow 0$ keeping $\Gamma$ constant.

%%%%%%%%%%%
\subsection{Structural quantities}

A key quantity to characterize the structure of the monolayer is the
radial distribution function (RDF) $g(r)$.
Considering a homogeneous isotropic distribution of domains in the monolayer plane, $g(r)$ represents the probability of finding a domain at the distance $r$ of another domain chosen as a reference point:
\begin{equation}
\label{gr}
     g(r)=\frac{1}{\rho} \left\langle \frac{1}{N} \sum_{\substack{i,j=1 \\i\neq j}}^N \delta({\bf r}-{\bf r}_i+{\bf r}_j) \right\rangle \; .
\end{equation}
Here, $\delta({\bf r})$ is the Dirac delta function and the angular brackets indicate an equilibrium ensemble average.

Another quantity to characterize structure properties is the static structure factor, $S(q)$ in dependence on the (scattering) wave number $\bf q$:
\begin{equation}
\label{Sq}
     S(q)=\left\langle \frac{1}{N} \sum_{\substack{i,j=1 }}^N \exp{({-i\bf q}\cdot[{\bf r}_i-{\bf r}_j])} \right\rangle \; .
\end{equation}
In 2D these two quantities are related to each other by
\begin{equation}
%\label{}
     g(r)=1+\frac{1}{2 \pi \rho} \int_0^\infty [S(q)-1] q J_0(qr) \, \mathrm{d}q.
\end{equation}
where $J_0(x)$ is the zeroth--order Bessel function of the first kind~\cite{lado1971}.

\section{Methods}

%%%%%%%%%%%%%%%%%%%%%%%%%%%%%%%%%%%%%
\subsection{Ornstein--Zernike}
\label{Sec:OZ}

For homogeneous and isotropic fluid whose particles interact through a pair potential, the Ornstein--Zernike (OZ) relation~\cite{hansen,ornstein1914} is defined as 
\begin{equation}
\label{eqOZ}
    h(r)=c(r)+\rho \int c(|\mathbf{r}-\mathbf{r'}|)h(r')\mathrm{d}\mathbf{r'},
\end{equation}
where $c(r)$ denotes the direct correlation function and 
\begin{equation}
    h(r)=g(r)-1,
\end{equation}
denotes the total correlation function.
Another equation that relates $h(r)$ to $c(r)$ is the exact non--linear closure condition
\begin{equation}
    g(r)=\exp[-\beta u(r)+h(r)-c(r)+B(r)],
\end{equation}
where $B(r)$ is the bridge function.
To close the system of equations, a formal exact expression for the bridge function in terms of the correlation functions is needed.

Some useful functions used in this formalism are the indirect correlation function
\begin{equation}
\label{eqgamma}
    \gamma(r)=h(r)-c(r),
\end{equation}
and the cavity distribution function~\cite{hansen}
\begin{equation}
    y(r)=\exp[\beta u(r)] g(r).
\end{equation}
Both functions, $\gamma(r)$ and $y(r)$, have the advantage of being continuous functions of $r$ even when there are discontinuities in $u(r)$ and hence in $g(r)$.
In terms of these functions, the bridge function takes the form:
\begin{equation}
\label{eqB}
    B(r)=\ln[y(r)] - \gamma(r).
\end{equation}

In default of an exact, closed form expression for $B(r)$, the bridge function is commonly approximated by a closure relation. We refer OZ-IET to as the closed integral equation system consisting of the OZ equation and a closure relation.
The simplest and most frequently used closure relations are the Percus--Yevick (PY)~\cite{hansen,percus1958} and Hypernetted--Chain (HNC)~\cite{hansen,vanleeuwen1959,meeron1960,morita1960a,morita1960b,rushbrooke1960,verlet1960,verlet1962} approximations.
The HNC closure consist in assuming  $B_{HNC}(r)=0$ while the PY  corresponds to setting  
\begin{equation}
\label{BPY}
B_{PY}(r) = \ln[1+\gamma(r)] - \gamma(r).
\end{equation}
The PY theory, in general, produces its best performances for short--ranged potentials, where it predicts with reasonable accuracy both structural and thermodynamic properties of hard sphere systems.
On the other hand, HNC works better for systems with long--range interaction potentials.

Apart from these two approximations, various specific forms of the bridge function or closure relations have been  proposed in order to improve the performance for different interaction potentials, such as Rogers--Young (RY)~\cite{rogers1984}, Verlet--modified (VM)~\cite{verlet1980}, Martynov--Sarkisov (MS)~\cite{martynov1983}, Balloni--Pastore--Galli--Gazillo (BPGG)~\cite{ballone1986}, Modified Hypernetted Chain (MHNC)~\cite{rosenfeld1979},  Reference Hypernetted Chain (RHNC)~\cite{lado1973,lado1982,lado1983}, Variational Modified Hypernetted Chain (VMHNC)~\cite{rosenfeld1986}, to name a few. 

For the particular case of the dipolar density potential in two dimensions, to the best of our knowledge, the integral equation theory has not been used to study this system.  

In this work, we selected the following closures to study their performance for this system.

\subsubsection*{Rogers--Young}

The Rogers and Young closure mixes PY closure at short distances and HNC closure at large distances by defining a mixing function $f(r)$. 
The resulting bridge function is given by~\cite{rogers1984}
\begin{equation}
   B_{RY}(r)=\ln\left[1+\frac{\exp[\gamma(r)f(r)]-1}{f(r)}\right]-\gamma(r),  
\end{equation}
where $f(r)=1-\exp[-\alpha r]$, and $\alpha$ is an adjustable parameter used  to force consistency between compressibility and virial equations of state
 (see Appendix~\ref{consist}).

\subsubsection*{MHNC}

The Modified Hypernetted Chain (MHNC) closure~\cite{rosenfeld1979} proposes to use a parameterized family of bridge functions taken from a known reference system, invoking the quasi--universality of the bridge functions. 

In three dimensions the hard spheres system (HS) has been extensively studied and different analytical expressions for the correlation functions
%g(r) + c(r)
(from which the bridge function can be obtained) are available, either within the PY approximation~\cite{wertheim1963,thiele1963} or from  phenomenological parameterization of the ``exact'' simulation data~\cite{verlet1972, henderson1975, carnahan1969}.
For this reason, the HS system is the usual choice for the reference system. 
Besides, the HS system has also the advantage of having only one parameter that determines the thermodynamic states, namely, the packing fraction $\phi$.
With this reference system, the bridge function results
\begin{equation}
\label{Bmhnc}
    B_{MHNC}(r)=B_{ref}(r;\phi_{eff}),
\end{equation}
where the parameter $\phi_{eff}$ is selected by requiring consistency between the equation of state obtained from the virial and the compressibility routes~\cite{rosenfeld1979, rosenfeld1986} (see Appendix~\ref{consist}). 

In two--dimensional systems, there is no analytical solution for the PY approximation of hard disks (PYHD). 
However, Adda-Bedia and coworkers~\cite{adda-bedia2008} developed a semi--analytic method to solve the PYHD equation and numerically computed the first 20 virial coefficients from the virial and compressibility routes of the equation of state.
For the purpose of this work, analytic expressions for the correlation functions (to compute the bridge function) are desired, since they need to be evaluated within the iterative solution of the OZ equation.
Different approximate expressions for $c_{PY}(r)$ or $g_{PY}(r)$ have been proposed in the literature~\cite{leutheusser1986,baus1986,baus1987,gonzales1991,yuste1993,ripoll1995,guo2006}.
Recently, Mier-y-Ter\'an \emph{et al.}~\cite{mier2018} compared three of these approximations (Refs.~\onlinecite{leutheusser1986,baus1987,gonzales1991}) and have shown that the Baus and Colot  Ansatz for $c_{PY}(r)$ gives the closest approximation to the PY structure, measured by its radial distribution function. 
With the aim of using the PYHD as reference system, we computed the PY bridge function based on the Baus and Colot Ansatz, but using the numerically computed PY virial coefficients obtained by Adda-Bedia  \emph{et al.}  (see Appendix~\ref{HDref}).
We will refer to this implementation of the MHNC closure as MHNC-PY.

On the other hand, analogous to the HS VWHG parameterizations~\cite{verlet1972,henderson1975}, Law and Buzza~\cite{law2009} suggested an extension of this approach to parameterize the correlation functions of hard disks.
Note, however, that the $g(r)$ resulting from these parameterizations do not fit simulation data as in the 3D Verlet--Weis approximation and other extensions~\cite{gonzales1991}.
%\
Using these expressions we computed the ``exact'' HD bridge function of the reference system (see Appendix~\ref{HDref} for details). 
This implementation of the MHNC closure will be referred to as MHNC-LB. 

%the bridge function is obtained from Eq.~(eqB) by using (specific) parameterizations for $y(r)$ and $\gamma(r)$ \cite{77, 78, 79}.

\subsubsection*{VMHNC}

The Variational Modified Hypernetted Chain (VMHNC) closure~\cite{rosenfeld1986} differs from the MHNC closure only in the procedure to determine the parameters of the reference system. 
 
In this scheme, and using HD as the reference system, the $\phi_{eff}$ is obtained by minimizing the VMHNC free energy functional~\cite{rosenfeld1986}. It can be shown that the requested extremum condition is satisfied when~\cite{rosenfeld1986} 
\begin{equation}
    \label{eqvar}
    \frac{d\delta(\phi_{eff})}{d\phi} - \frac{\rho}{2} \int [g(r) - g_{ref}(r;\phi_{eff})] \frac{\partial B_{ref}(r;\phi_{eff})}{\partial \phi} d\mathrm{r} = 0.
\end{equation}
\noindent
Here, $\delta(\phi)$ is a fitting function to improve 
the VMHNC approach~\cite{rosenfeld1986}.
For the particular case of the 3D PY hard sphere reference system, Rosenfeld~\cite{rosenfeld1986} obtained an accurate estimate of $\delta(\phi)$, given by the simple expression
\begin{equation}
    \label{delta}
    \delta(\phi)=f_{CS}(\phi) - f_{PYHSv}(\phi), 
\end{equation}
where $f_{CS}(\phi)$ and $f_{PYHSv}(\phi)$ are the empirical Carnahan--Starling~\cite{carnahan1969} free energy and the Percus--Yevick virial free energy, respectively.
Rosenfeld~\cite{rosenfeld1986} further suggests using this fitting function, Eq.~(\ref{delta}), for any interaction potential.
For two--dimensional systems (2D), we extended the previous result using the
accurate equation of state proposed by Santos \emph{et al}. \cite{santos1995} and the PY virial coefficients computed by Adda--Beddia \emph{et al}.~\cite{adda-bedia2008}.
Following Rosenfeld~\cite{rosenfeld1986}, we define the fitting function as
\begin{equation}
    \label{delta2d}
    \delta(\phi)=f_{S}(\phi) - f_{PYHDv}(\phi), 
\end{equation}
where $f_{S}(\phi)$ is the Helmholtz free energy obtained from the empirical Santos~\cite{santos1995} equation of state, and the Percus--Yevick virial free energy, $f_{PYHDv}(\phi)$, is calculated using the 20 first virial coefficient obtained by Adda-Bedia~\cite{adda-bedia2008}.

The scheme resulting from Eqs.~(\ref{eqvar}) and (\ref{delta2d}) will be referred as VMHNC-PY.

If instead of using the PY approximations for the HD reference system,  the parameterized ``exact'' HD expressions are used, this closure reduces to Eq.~(\ref{eqvar}) with $\delta(\phi)=0$, and will be referred as VMHNC-LB.
Note that this scheme is also known in the literature as  Reference Hypernetted Chain (RHNC)~\cite{lado1982,lado1983,castello2021}.

%%%%%%%%%%%
\subsection{Simulations and numerical methods}

We compare the results of the different OZ--IETs with Metropolis Monte Carlo (MC) simulations.
The simulated systems consisted of $N=1024$ disks of radius $R$ under periodic boundary conditions, using the minimum image convention. The size of the simulation box, $L$, was determined using the expression of the condensed area fraction $\phi = N\pi R^2 /L^2$.
The disks interact with each other under a dipolar density pair potential. 
For the typical experimental parameters range, the first coordination shell is located outside the region where any of the asymptotic expressions, Eq.~(\ref{Ud_as}), are valid. 
Therefore, we work with the full expression of the potential, Eq.~(\ref{Ud}). In order to compute it, the 4D--integral is reduced to a single integral that involves an elliptic function, as it is shown in Appendix~\ref{integral_density}). 
The energies are calculated using Eq.~(\ref{Ud_3}). 
To update the position of each disk (randomly chosen), we randomly set a 
trial 2d--displacement, accepted according to Metropolis rules.
To compute the thermal averages of the structural observables, first, we run $4 \times 10^5$ Monte Carlo Steps (MCS) to thermalize and then use $8 \times 10^5$ MCS to measure the quantities, computing these quantities every $500$ MCS and averaging them.

The OZ integral equation is numerically solved using the Ng fast--converging iteration scheme~\cite{ng1974} with 5 parameters. 
Fourier transformations are computed using the Lado algorithm~\cite{lado1971}, which imposes a discretization of the $r$--space given by the roots of the zeroth--order Bessel function of the first kind, $J_0(x)$, scaled to cover the interval $[0,r_{cut}]$, with the cutoff distance, $r_{cut}$, sufficiently large to assume that the integrals (and transforms) may be truncated at $r_{cut}$.
Integrals are performed using the trapezoidal rule (with unequal intervals), taking into account discontinuities of the integrands by splitting the integration interval and extrapolating to obtain the values at the discontinuity.
For improving the convergence of the iteration scheme, the number density of the system was linearly increased (using between 10 to 40 steps) from zero to the desired value, solving the OZ equation at each density using the solution of the previous density as the initial guess.
In the present work, $N=8000$ discretization points were used, and the cutoff distance, $r_{cut} = 45\times r_{m}$, was chosen. Here, $r_{m}$ represents the mean geometrical distance.

%%%%%%%%%%%%%%%%%%%%%%
\section{Results and Discussion}
%%%%%%%%%%%%%%%%%%%%%%

We now illustrate the results of the different closures described above; RY, MHNC--PY, MHNC--LB, VMHNC--PY and VMHNC--LB. We compare them with MC simulations for a set of state points that both, belong to the fluid region of the  $\phi$--$\Gamma$ phase diagram, and the interaction is strong enough that the disks do not come into contact.
The performance of these closures is assessed in terms of the radial distribution function and the structure factor.
\begin{figure*}[h!]
\centerline{
\includegraphics[width=1.03\textwidth,angle=0]{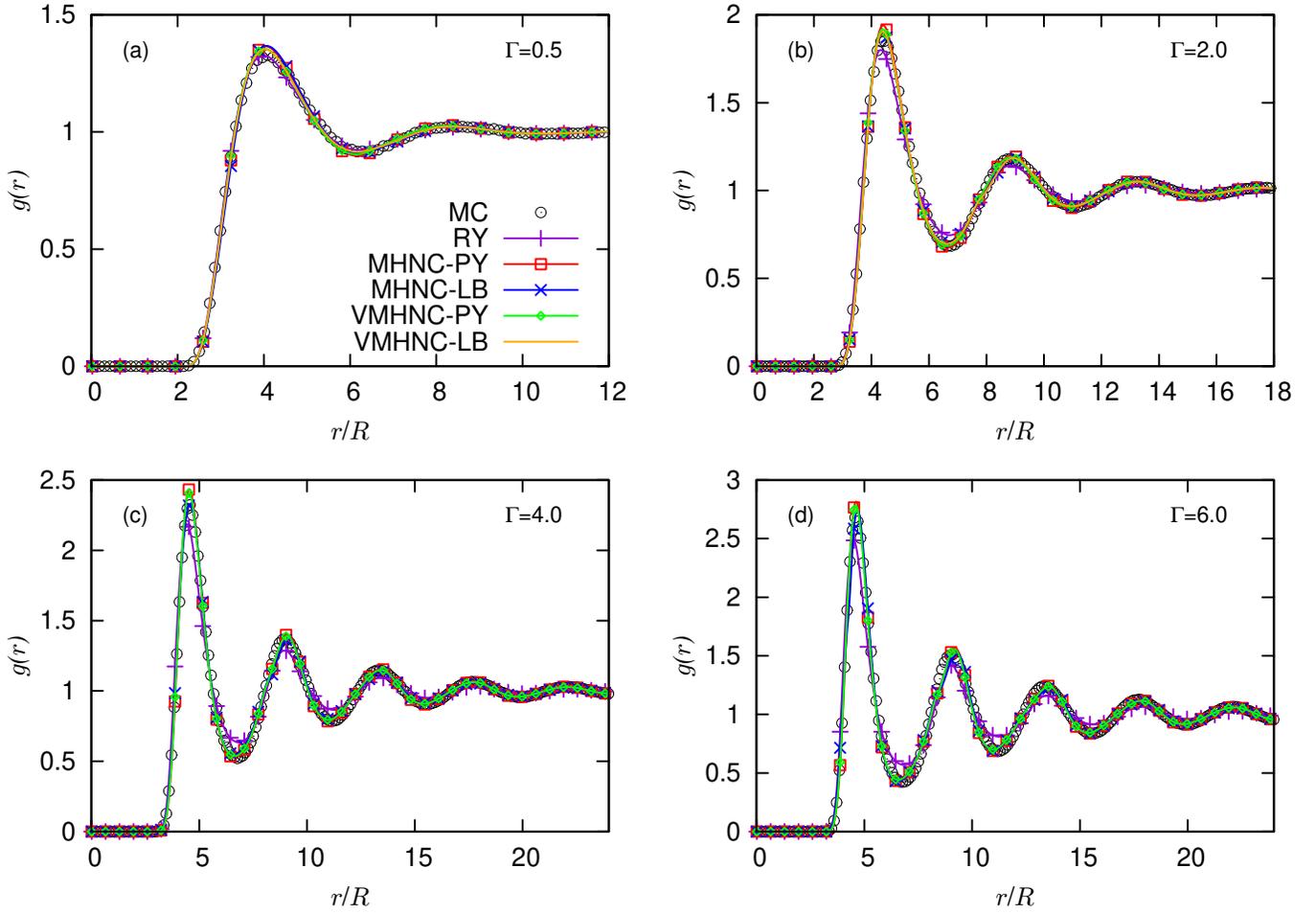}}
\caption{Radial distribution functions for $\phi=0.15$ and $\Gamma=0.5, 2.0, 4.0$ and $6.0$; panels (a), (b), (c) and (d), respectively. Circles correspond to MC simulations and solid lines to RY (violet), MHNC--PY (red), MHNC--LB (blue), VMHNC--PY (green) and VMHNC--LB (orange) closures. The symbols in the IETs results were plotted at arbitrary data intervals as a guide to the eye.}
\label{gr_phi0.15}
\end{figure*}
%

%%%%%%%%%%%%%%%%%%%%%%%%%%%%%%%%%%%%%%%%%%%%
\subsection{Radial distribution function}

The calculated radial distribution functions, $g(r)$, are shown in Fig.~\ref{gr_phi0.15} for $\phi=0.15$ and $\Gamma=0.5, 2.0, 4.0$ and $6.0$.

For the less structured system, $\Gamma=0.5$, all closures reproduce globally well the simulations results as shown in panel (a).
There are slightly differences in the performance of the studied closures only around the first peak, where the RY approach performs very well.

For larger values of the coupling parameter, $\Gamma$, the RY approach starts to deviate from the MC results.
Already at $\Gamma=2$ (panel (b)), there are appreciable differences, not only at the first peak but also at the first minimum and second peak.
The performance becomes even worse for more structured systems (panels (c) and (d)), showing large discrepancies with the MC data around peaks and minima. 

The results obtained with the MHNC--LB closure are globally in good agreement with MC data.
On closer inspection, the agreement of the first peak height improves with increasing $\Gamma$. In particular, for $\Gamma = 4$ this closure results the best in comparison with the other closures under consideration.
Note that, as in other schemes based on VWHG parameterizations of the pair correlation and the cavity functions, a small unphysical shoulder appears on the left of the second peak~\cite{lado1983}. 

Similarly to the MHNC--LB, the VMHNC--LB closure presents good global performance, but only up to $\Gamma \approx 3.5 $.
For larger coupling parameter values, Eq~(\ref{eqvar}) can no longer be satisfied. 
This closure is also based on a VWHG--like parameterization, for this reason, the pair correlation function presents an unphysical shoulder.

In contrast to the closures based on a VWHG--like parameterizations, the schemes that use HDPY reference system, MHNC-PY and VMHNC-PY, have solution in all the studied parameter range.

MHNC--PY and VMHNC--PY results are globally in quite good agreement with the MC data for the studied systems (panels a--d), except for the first two peaks regions. There, they slightly overestimate the peak values systematically, with the VMHNC--PY closer to the MC results.

A qualitative difference between the MC results and the MHNC--LB, VMHNC--LB and VMHNC-PY approaches can be observed in the more structured systems, panels (c) and (d). 
There, the shape of the second and third peaks tends to lean to the right, which leads to the position of their maxima also shifting to the right.

Note that, for a given reference system, PY or LB, the radial distribution functions obtained using the MHNC and VMHNC closures are very similar. For PY--based closures the percentage difference around the first peak is $\approx 2\%$.
At this point, it is worth mentioning that the LB--based schemes are more numerically sensitive for quite structured systems. 
The numerical derivative of the \textit{virial} pressure (MHNC) and the bridge function (VMHNC) need to be carefully calculated, implying that the number of points in the discretization might need to be adjusted or the increment in the finite difference derivative carefully selected. 
For these schemes, we also found that within the range of parameters that we have considered, there is not always a solution, and eventually, close to where the solution is lost, more than one solution might exist (the one with a lower effective area fraction was selected).
The PY--based approaches, on the other hand, are numerically preferable, and for the VMHNC case having an analytical implementation of the derivative of the bridge function is of advantage.

Other systems in the liquid region were studied, giving similar results. Radial distribution functions for $\phi=0.05$ and $0.25$ and different $\Gamma$ values are shown in Appendix~\ref{phi005_025}.
%Supplementary Material \footnote{See Supplemental Material at [URL will be inserted by publisher] for complementary results for the pair correlation function and the structure factor for area fractions $\phi=0.05$ and $\phi=0.25$. }. 
There, we also consider a particular case where the contact value is different from zero.
\begin{figure*}[t]
\centerline{
\includegraphics[width=1.03\textwidth,angle=0]{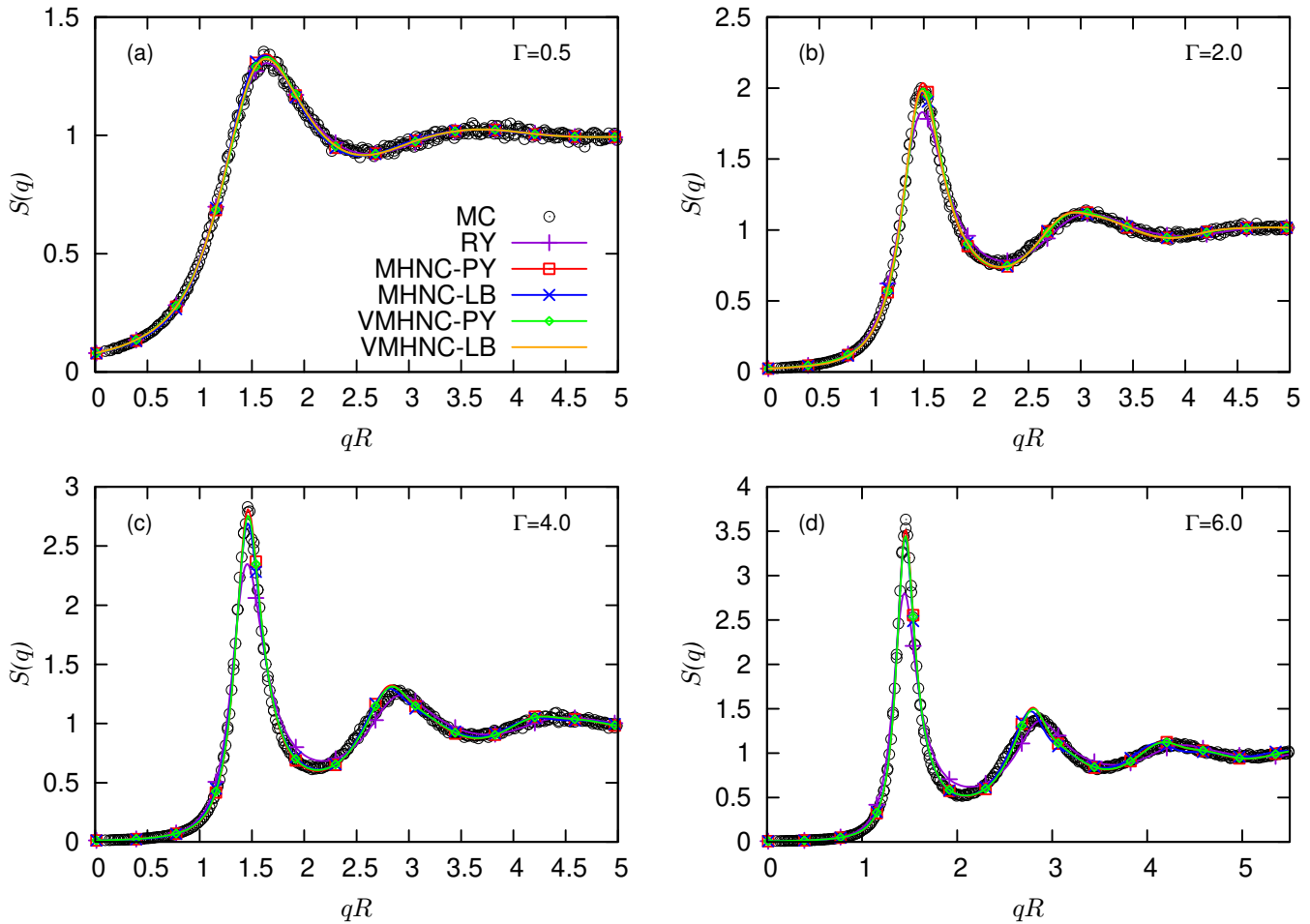}}
\caption{Structured factors for $\phi=0.15$ and $\Gamma=0.5, 2.0, 4.0$ and $6.0$; panels (a), (b), (c), and (d), respectively. Circles correspond to MC simulations and solid lines to RY (violet), MHNC--PY (red), MHNC--LB (blue), VMHNC--PY (green), and VMHNC--LB (orange) closures. The symbols in the IETs results were plotted at arbitrary data intervals as a guide to the eye.}
\label{sq_phi0.15}
\end{figure*}
%

%%%%%%%%%%%%%%%%%%%%%%%%%%%%%%%%%%%%%%%%%%%%
\subsection{Structure Factor}

The results obtained for the structure factor, $S(q)$, are presented in Fig.~\ref{sq_phi0.15} for the systems shown above.
We first note that the statistical error of the MC data for $S(q)$ is much larger than the obtained for the pair correlation function, as it is expected, since the structure factor can not be calculated averaging over all particles, as in the $g(r)$ computation.

In panel (a) of Fig.~\ref{sq_phi0.15}, we observe that all closures perform very well within the statistical error of the MC simulation data. 
At the first peak, the RY closure slightly underestimates the MC results, while the other closures are almost indistinguishable.

The same good agreement is observed for $\Gamma=2.0$ (panel (b)) except for the RY closure which already at this level of structure, $S(q_m)\approx 2$, has a poor performance. For more structured systems (see panels (c) and (d)), the RY performance continues to deteriorate.

On the other hand, for $\Gamma=2.0$ (panel (b)) the other closures under consideration are in relatively good agreement with the MC results. 
The first peak height is well captured by MHNC--LB, VMHNC--LB, and VMHNC--PY, while MHNC--PY slightly overestimates it. At the first minimum, these closures underestimate the MC data by a very small margin. A qualitative difference, even though slight, is observed in the second peak, where the peak position of these closures are at smaller $q$--values, and the shape is asymmetric with respect to its maximum.
%\tb{(like a right skewed shape). 
This \textit{skewed} shape is more noticeable than that found in $g(r)$.
For larger coupling parameter values (panels (c) and (d)), this difference becomes more pronounced, and the second peak height is overestimated. Analogous differences are visible in the third peak. At the same time, the first peak height tends to be slightly underestimated.

Analogous to the case of the $g(r)$, the results obtained using the MHNC and VMHNC closures are very similar. Furthermore, the PY--based closures have solutions for all the studied systems.

Similar results were found for $\phi=0.05$ and $0.25$, with different $\Gamma$ values, which are shown in  Appendix~\ref{phi005_025}.
%the Supplementary Material \cite{Note1}.

%%%%%%%%%%%
\section{Conclusions}
%%%%%%%%%%%
In this work, we investigated the performance of Ornstein--Zernike--based integral equation theories for the dipolar density interaction in 2D systems. We have studied three closures of the Ornstein--Zernike equation: Rogers--Young (RY), Modified Hypernetted Chain (MHNC), and Variational Modified Hypernetted Chain (VMHNC). 
Besides, two approximations of the hard disk reference system for these last two closures were considered; one based on the Percus--Yevick approximation (PY) and the other based on an extension of the hard spheres Verlet--Weis--Henderson--Grundke parameterization (LB).
The performance of each closure was evaluated by comparing the results with Monte Carlo simulations in terms of the radial distribution function and the structure factor. 

The results showed that, for the less structured system, all closures reproduce the simulations globally well, with only slight differences around the first peak, where the RY accurately reproduces the $g(r)$, but slightly underestimates the peak height of $S(q)$.
For more structured systems, the RY approach starts to deviate from the MC results, and already systems with $S(q_m)\approx 2$ ($g(r_m)\approx 2$) have a poor performance.
This is in accordance with the results obtained for point dipole interacting colloids in 2D by Hoffmann \textit{et al.}~\cite{hoffmann2006a}.
There, the RY closure performs very well for a system with $g(r_m)\approx 1.5$, and it presents a poor performance for the system with $g(r_m)\approx 2.6$.  

For more structured systems, the other closures under consideration -- MHNC--LB, MHNC--PY, VMHNC--LB, and VMHNC--PY -- are in relatively good agreement with the MC results.
There are four key points about these closures that should be highlighted.
First, up to a certain structure, the closures that use LB as a reference system perform well globally, but for larger coupling parameters values, the thermodynamic consistency (in the case of MHNC--LB) or Eq~(\ref{eqvar}) (in the case of VMHNC--LB) cannot be satisfied.
Second, the pair correlation function and the structure factor of the LB-based closures exhibit an unphysical shoulder in the second peak, as expected for any closures based on a VWHG--like parameterization.
Third, for a given reference system, PY or LB, $g(r)$ and $S(q)$ obtained using the MHNC and VMHNC closures are fairly similar.
Fourth, in the more structured system under consideration, there is a qualitative difference, even though slight, between the MC results and these closures; the second and third peaks tend to have a rightward lean in the case of $g(r)$ and a leftward lean in the case of $S(q)$. These qualitative differences are expected to worsen for even more structured systems.

On closer examination, when measured by the pair correlation function, it becomes clear that LB-based closures are preferable to PY ones. 
However, when the structure factor is employed as a measure, the reference system that performs better depends on the specific region of the phase diagram under consideration.

These results show that, at least, in a great part of the phase diagram of dipolar density interacting disks monolayers the MHNC and VMHNC OZ--IETs perform quite accurately, becoming an interesting tool for systematic studies of this system or for producing structural data needed as input for other theories.

%%%%%%%%%%%%%%%%%%%%%%%% Acknowledgments  %%%%%%%%%%%%%%%%%
\section*{Acknowledgments}
The authors acknowledge financial support from Fondo para la Investigaci\'on Cient\'ifica y Tecnol\'ogica, Argentina (FonCyT) under grants  PICT2015-0735 and PICT2020--SerieA--02931,  Secretar\'ia de Ciencia y T\'ecnica de la Universidad Nacional de C\'ordoba, Argentina (SECyT--UNC) under grant No. 33620180100018CB.
The authors thank Marco Heinen for valuable discussions at the beginning of the project, and Martin Buzza for helpful discussions regarding the parameterized hard disk correlation functions (Ref.~\onlinecite{law2009}). ERF acknowledges support from the International Center of Theoretical Physics (ICTP) through the Associates Programme (2022-2027).

\vspace*{1cm}

%%%%%%%%%%%%%%%%%%%%%%%%    Apendices 
\appendix

\section{Pair correlation function and structure factor for area fractions $\phi=0.05$ and $\phi=0.25$}
\label{phi005_025}

\begin{figure*}[h!]
\centerline{
\includegraphics[width=1.03\textwidth,angle=0]{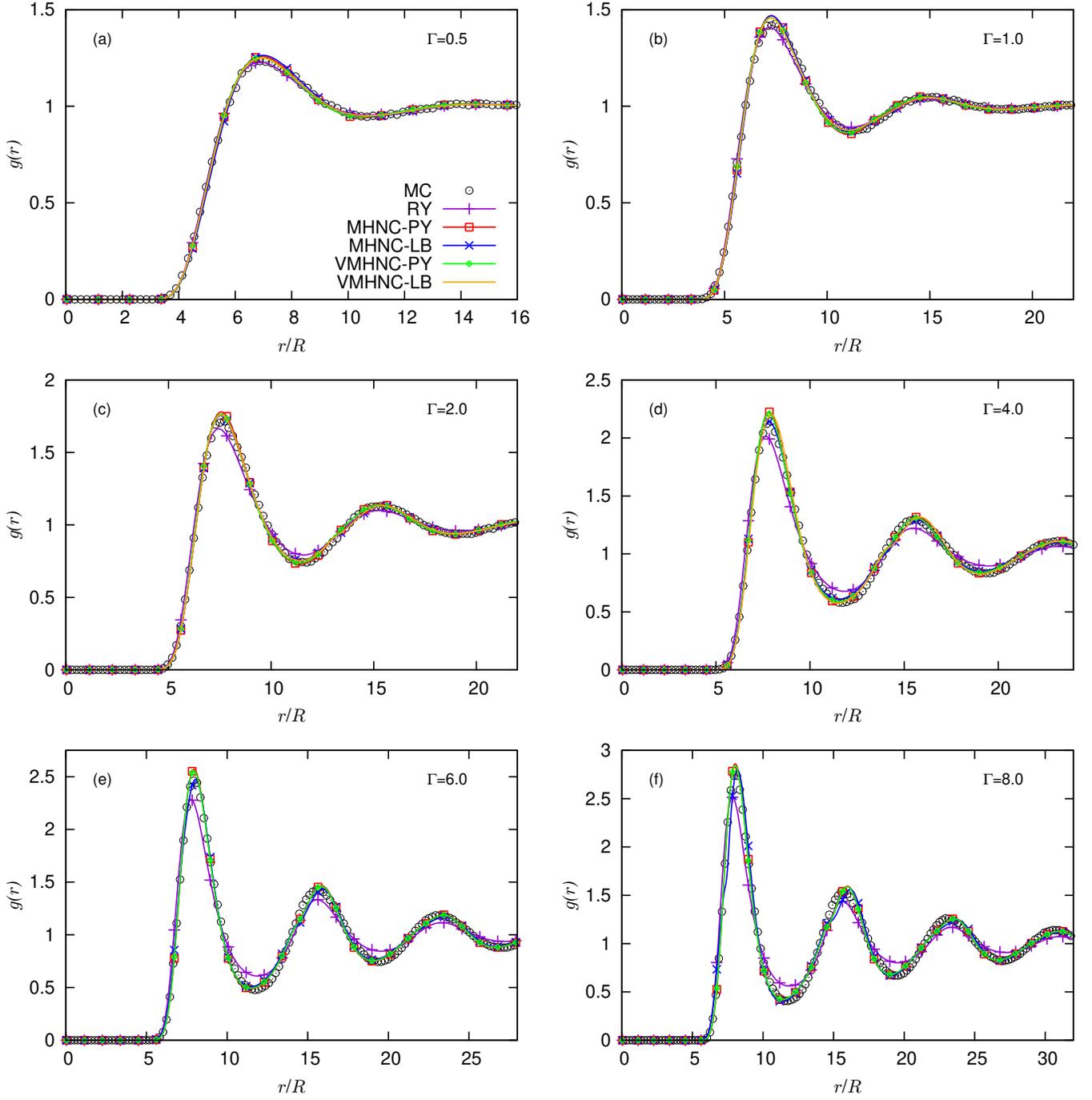}}
\caption{Radial distribution functions for $\phi=0.05$ and $\Gamma=0.5, 1.0, 2.0, 4.0, 6.0$ and $8.0$; panels (a), (b), (c), (d), (e) and (f), respectively. Circles correspond to MC simulations and solid lines to RY (violet), MHNC--PY (red), MHNC--LB (blue), VMHNC--PY (green) and VMHNC--LB (orange) closures. The symbols in the IETs results were plotted at arbitrary data intervals as a guide to the eye. }
\label{gr_phi0.05}
\end{figure*}

\begin{figure*}[h!]
\centerline{
\includegraphics[width=1.03\textwidth,angle=0]{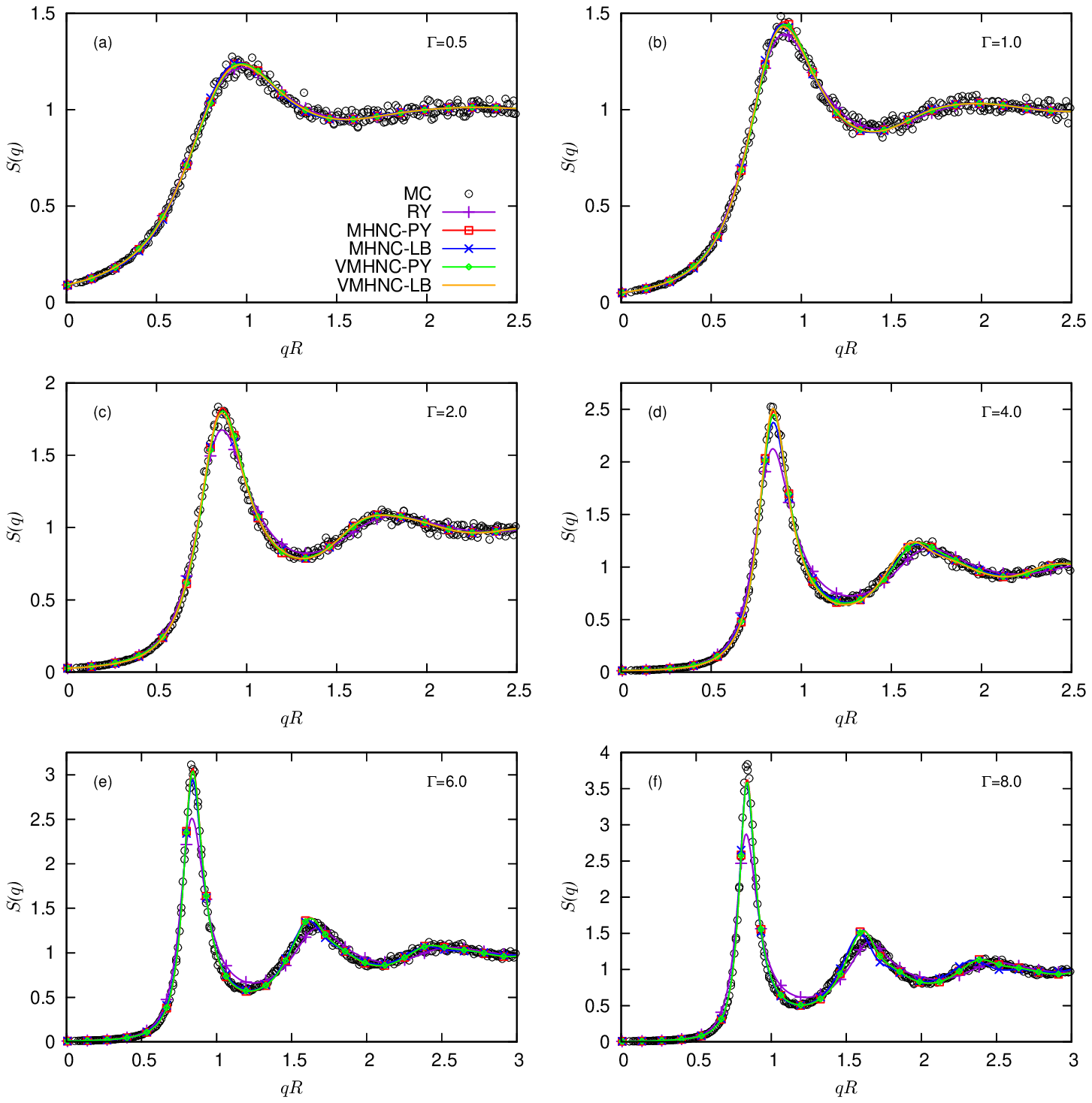}}
\caption{Structured factors for $\phi=0.05$ and $\Gamma=0.5, 1.0, 2.0, 4.0, 6.0$ and $8.0$; panels (a), (b), (c), (d), (e) and (f), respectively. Circles correspond to MC simulations and solid lines to RY (violet), MHNC--PY (red), MHNC--LB (blue), VMHNC--PY (green) and VMHNC--LB (orange) closures. The symbols in the IETs results were plotted at arbitrary data intervals as a guide to the eye.}
\label{sq_phi0.05}
\end{figure*}

\begin{figure*}[h!]
\centerline{
\includegraphics[width=1.03\textwidth,angle=0]{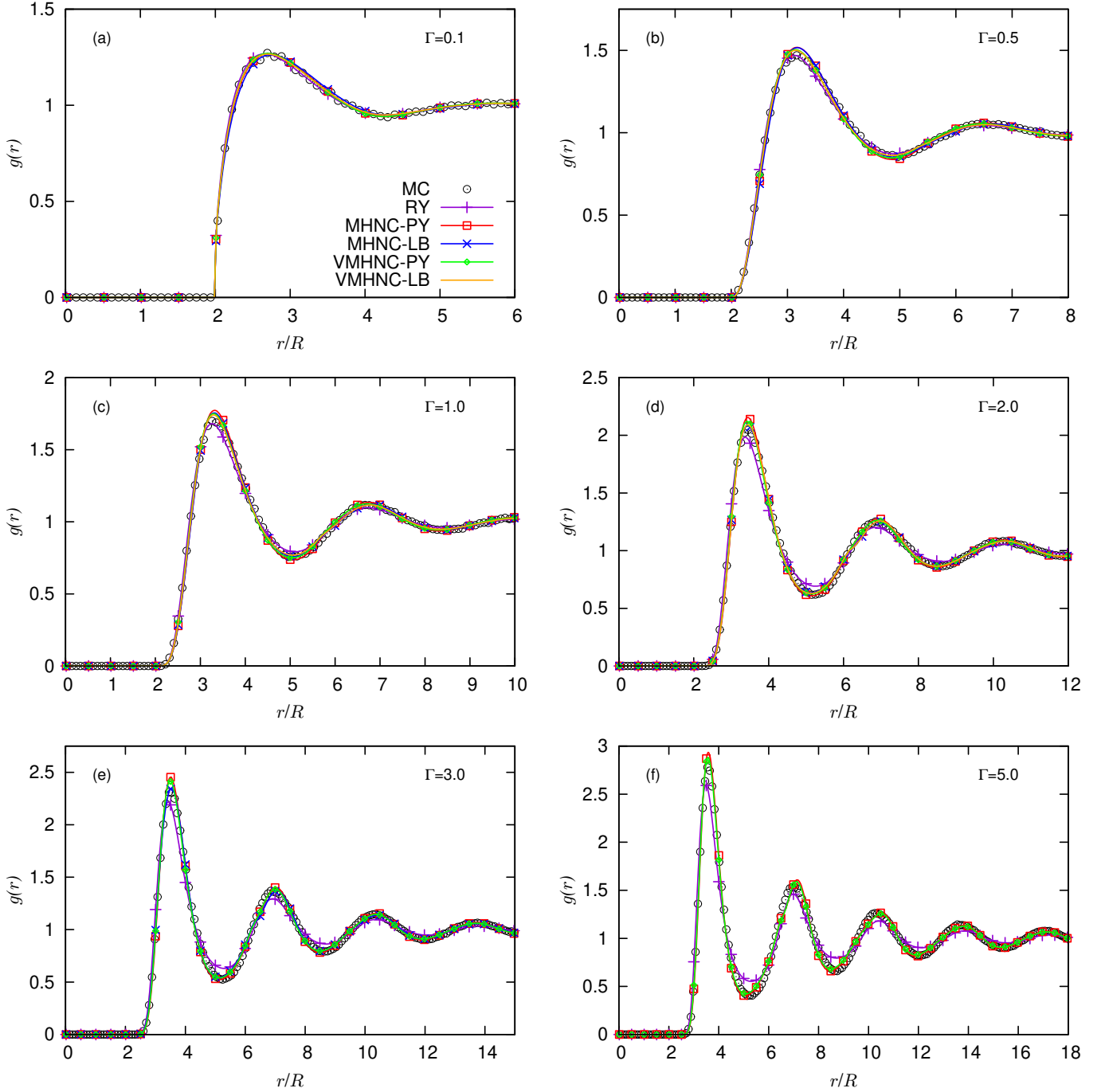}}
\caption{Radial distribution functions for $\phi=0.25$ and $\Gamma=0.1,0.5, 1.0, 2.0, 3.0$ and $5.0$; panels (a), (b), (c), (d), (e) and (f), respectively. Circles correspond to MC simulations and solid lines to RY (violet), MHNC--PY (red), MHNC--LB (blue), VMHNC--PY (green) and VMHNC--LB (orange) closures. The symbols in the IETs results were plotted at arbitrary data intervals as a guide to the eye. Panel (a) shows that the closures maintain their excellent performance even when the hard core potential is relevant, i.e., the contact value is different from zero. }
\label{gr_phi0.25}
\end{figure*}

\begin{figure*}[h!]
\centerline{
\includegraphics[width=1.03\textwidth,angle=0]{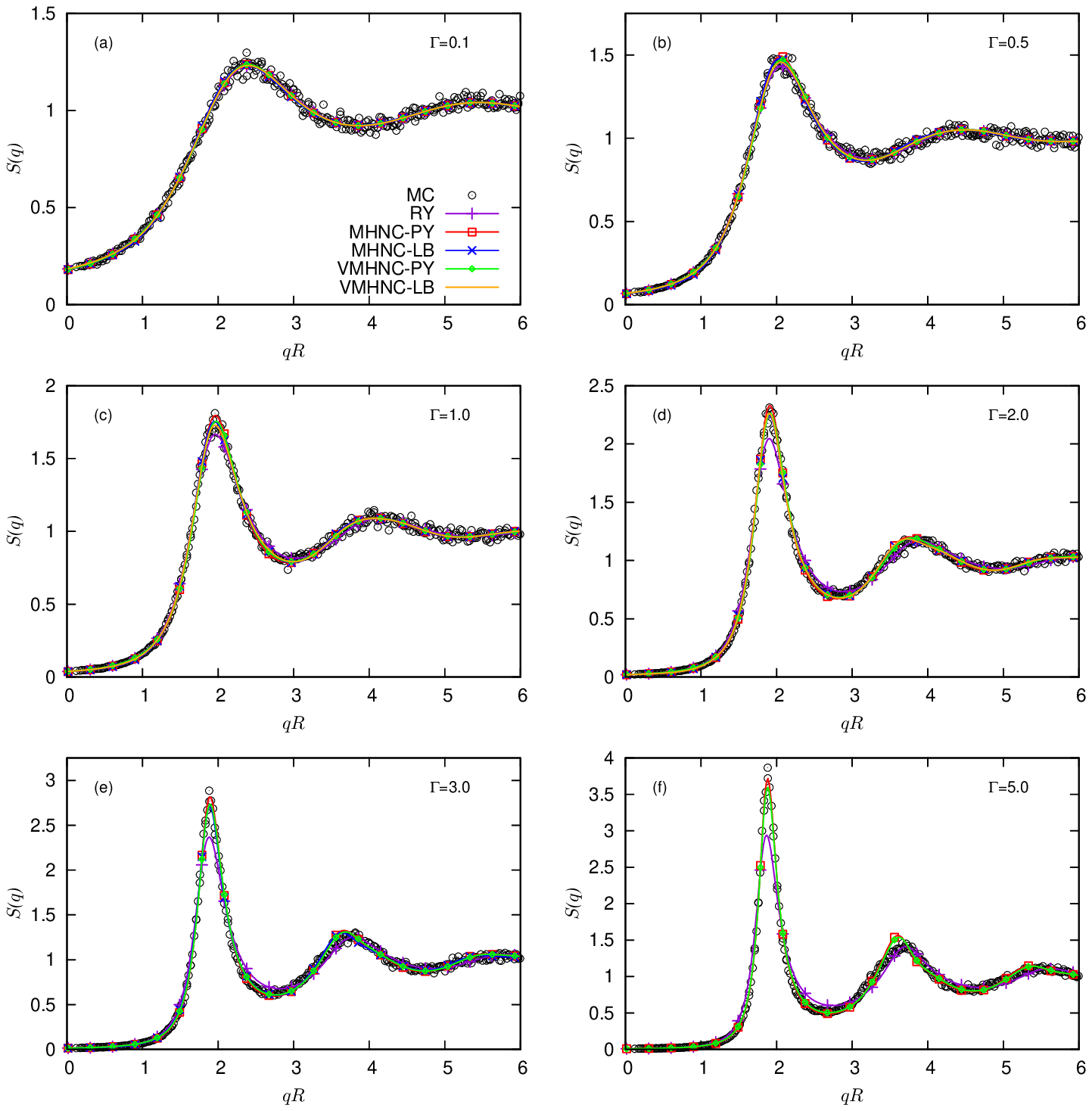}}
\caption{Structured factors for $\phi=0.25$ and $\Gamma=0.1, 0.5, 1.0, 2.0, 3.0$ and $5.0$; panels (a), (b), (c), (d), (e) and (f), respectively. Circles correspond to MC simulations and solid lines to RY (violet), MHNC--PY (red), MHNC--LB (blue), VMHNC--PY (green) and VMHNC--LB (orange) closures. The symbols in the IETs results were plotted at arbitrary data intervals as a guide to the eye. From panel (a) it can be observed that the closures continue to perform very good even when the hard core potential is relevant.}
\label{sq_phi0.25}
\end{figure*}

%\subsection{Area fraction $\phi=0.05$}
%\subsubsection{Pair correlation function}
%
Results for the pair correlation function and the structure factor for $\phi=0.05$ are shown in figures~\ref{gr_phi0.05} and \ref{sq_phi0.05}, respectively.
For the area fraction $\phi=0.25$ the corresponding results are shown in figures~\ref{gr_phi0.25} and \ref{sq_phi0.25}.

%
%
%\newpage
%\subsubsection{Structure Factor}
%
%
%
%\newpage
%subsection{Area fraction $\phi=0.25$}
%\subsubsection{Pair correlation function}
%
%Figures~\ref{gr_phi0.25} and \ref{sq_phi0.25}, show results for the pair correlation function and the structure factor, respectively.
%
%
%
%\newpage
%\subsubsection{Structure Factor}
%
%

%%%%%%%%%%%%%%%%%%% end -- old Supp. Mat.   %%%%%%%%%%%%%%%%%%%%%%%%
%\counterwithin*{equation}{section}
%\renewcommand{\theequation}{\thesection.\arabic{equation}}

\section{Dipolar density integral}
\label{integral_density}

The dipolar density energy is given by Eq. (\ref{Ud}).
For monodisperse systems (all domain radii equal to $R$), Wurlitzer \textit{et al.}~\cite{fischer02} found that it can be expressed in terms of a single integral:
\begin{equation}
\label{Ud_3}
 U_d(r)=f_0 \int_{0}^{2R}\frac{4r'}{(r'+r)(r'-r)^2} \; E\left(\frac{4r'r}{(r'+r)^2}\right) p(r')\;\mathrm{d}r',
\end{equation}
where $E(q)=\int_0^{\pi/2}\sqrt{1-q\sin^2{\theta}}\;\mathrm{d}\theta$ is the complete elliptic integral of second kind and 
\begin{equation}
\label{p2}
 p(r)=-r\sqrt{R^2-(r/2)^2} +2R^2 \left[\frac{\pi}{2}-\arcsin\left(\frac{r}{2R}\right)\right].
\end{equation}
In order to obtain the potential constant $C$ appearing in Eq.~(\ref{Ud_as}), we make the numerical integration of eq. (\ref{Ud_3}) with $r=2R$.

\section{Thermodynamic inconsistencies in the OZ--IET formalism}
\label{consist}

As a consequence of the absence of exact expressions for the bridge function, or equivalently the use of approximate closure relations, inconsistencies appear in the thermodynamic quantities calculated within the OZ--IET formalism (OZ + closure).
The equation of state (EOS) obtained from different ``routes'' %(\emph{virial, compressibility} and \emph{energy}) 
in general will differ. The most common routes, particularized for a 2--dimensional system, are:  
\begin{itemize}
\item[]{\textit{Virial route }} 

The \textit{virial} compressibility factor is given by
\begin{eqnarray}
    Z_v(\rho,\beta) &\equiv&
    \frac{\beta P_v}{\rho} \nonumber \\ 
    &=& 1 - \frac{\pi \beta \rho}{2} 
    \int_0^\infty 
    g(r)\frac{\partial u(r)}{\partial r} r^2  \;\mathrm{d}r  \; ,
    \label{vEOS}
\end{eqnarray}
\noindent
Here, $u(r)$ is the inter--particle pair potential.
The corresponding \textit{virial} EOS is then straightforwardly obtained. The subscript $v$ indicates that the corresponding quantity is obtained through the \textit{virial} route.
%La EOS es despejar P: P=rho/beta (1-...)

\item[]{\textit{Compressibility route }}

From the compressibility equation, the thermal compressibility can be written as
\begin{eqnarray}
   \chi_c(\rho,\beta) &\equiv& 
   \left( \beta \frac{\partial P_c}{\partial \rho}\right)^{-1}_T \nonumber \\ 
    &=& 1 + 2 \pi \rho
   \int_0^\infty h(r) r \;\mathrm{d}r \;.
    \label{cEOS}
\end{eqnarray}
\noindent 
%\tb{OJO! $\chi^{-1} = (\beta/\rho) \, \chi^{-1}_{Hansen}$} \\
%
\noindent
Then, the compressibility factor  $Z_c(\rho,\beta)$ can be determined by integrating $1/\chi_c$ with respect to the density and along an isothermal path. This determines the \textit{compressibility} EOS.
In Eq.~(\ref{cEOS}) the subscripts $c$ emphasizes that it is obtained through the \textit{compressibility} route.
Invoking the OZ equation, the compressibility equation can be conveniently written in terms of $c(r)$ as
\begin{equation}
    \chi_c^{-1}(\rho,\beta)
   =  1 - 2\pi \rho  \int_0^\infty 
   c(r) r \;\mathrm{d}r \; . \label{chim1_c}
\end{equation}

\item[]{\textit{Energy route }} 

The internal energy per particle is given by
\begin{equation}
    \frac{U}{N} = \frac{1}{\beta} 
    + \pi \rho \int_0^\infty g(r) u(r) r \;\mathrm{d}r \; . \label{u}
\end{equation}
\noindent 
Since,
\begin{equation}
    U = \left( \frac{\partial \beta F}{\partial \beta} \right)_V \; ,
\end{equation}
\noindent
the Helmholtz free energy, $F$, can be obtained by integrating $U$ with respect to the inverse temperature $\beta$ along an isochore path. Here, $V$ denotes the area.
Then, the pressure $P_e$, and consequently the \textit{energy} EOS, is obtained as
\begin{equation}
    P_e = - \left( \frac{\partial F}{\partial V}\right)_T \; .
\end{equation}
\noindent
From which the \textit{energy} compressibility factor, $Z_e(\rho,\beta)$, is straightforwardly obtained.
The subscript $e$ indicates that the corresponding quantity is obtained through the \textit{energy} route.

\end{itemize}The subscript $e$ indicates that the corresponding quantity is obtained through the \textit{energy} route.

The different ways of obtaining the compressibility factor (or the EOS) are equivalent if the exact correlation functions for the system are used.
However, as mentioned above, for approximate correlation functions, in general they are not, and so they are thermodynamically inconsistent.

In practice, the self--consistency conditions are written as two different equations, namely,
\begin{eqnarray}
\left( \beta \frac{\partial P_v}{\partial \rho} \right)_T
&=& \chi_c^{-1}(\rho,\beta) \; ,  \label{v-c} %\textbox{virial-compressibility}
\\
\left( \frac{1}{\rho} \frac{\partial \beta P_v}{\partial \beta} \right)_\rho &=& 
\frac{1}{V} \left( \frac{\partial U}{\partial \rho} \right)_T \; . \label{v-e} %\textbox{virial-energy}
\end{eqnarray}
\noindent
Here, the left--hand sides are meant to be computed using Eq.~(\ref{vEOS}), while the right--hand sides using Eq.~(\ref{chim1_c}) and Eq.~(\ref{u}), respectively.
The advantage of these equations with respect to the equations for the compressibility factors is that one can verify the consistency without the necessity to perform integrals over ``paths'' of thermodynamic states. 

The virial--compressibility consistency equation, Eq.~(\ref{v-c}), is used in our implementations of the RY and MHNC schemes.
Note also that for the HNC and VMHNC closures, the virial route is consistent with the energy route~\cite{morita1960b,rosenfeld1986,lado1982}, i.e., Eq.~(\ref{v-e}) is verified.

\section{Hard Disk reference systems}
\label{HDref}

The MHNC and VMHNC schemes, presented in Sec.~\ref{Sec:OZ}, are based on two parameterized families of bridge functions (and RDFs in case of VMHNC) from the HD reference system.

The different reference system bridge functions (and RDFs) used here are obtained from either Percus--Yevick Hard Disk (PY) or from parameterizations from ``exact'' hard disk (LB) approaches. 
Note that in contrast to the hard spheres case, in 2D the PY solution needs to be approximated, and there is no unique widely used parameterization of the ``exact'' hard disk correlation functions or EOS.

\subsection{PY: PY hard disks}
\label{subsec:PY}

To approximate $c_{PYHD}(r)$ we have slightly modified the Baus and Colot~\cite{baus1986,baus1987} semi--empirical expression 
by computing the expansion coefficients, $c_n$, in the rescaled  and truncated  virial series of the compressibility factor, 
\begin{equation}
    Z_N(\phi) = \frac{1 + \sum_{n=1}^{N} c_n \, \; , \phi^n}{(1-\phi)^2} \label{ZN}
\end{equation}
\noindent
using the virial coefficients from the compressibility route obtained by Adda--Bedia \textit{et al.} and truncating the series at $N=19$.
The coefficients $c_n$ are obtained from the PY--virial coefficients $B_i$~\cite{adda-bedia2008} (from the compressibility route) using Eq.~(2.4) and Eqs.~(2.15--2.17) from Ref.~\onlinecite{baus1987}.

The explicit form for the direct correlation function results~\cite{baus1987} in:
\begin{eqnarray}
    c_{PYHD}(x;\phi)&=&-\frac{\partial}{\partial\phi}[\phi Z_N(\phi)]\Theta(1-x)\nonumber\\
    && \times \left[ 1-a^2\phi+a^2\phi\; \omega(x/a) \right],
\end{eqnarray}
where $x=r/(2R)$ and $\beta=k_B T$. 
The function $\omega(x)$ is 
\begin{equation}
    \omega(x)=\frac{2}{\pi}[\arccos(x)-x\sqrt{1-x^2}],
\end{equation}
and $a=a(\phi)$ is a scaling function that can be numerically obtained from the following algebraic equation:
\begin{eqnarray}
\frac{2}{\pi} \left[ a^2(a^2-4)\arcsin(1/a)-(a^2+2)\sqrt{a^2-1} \right] \nonumber\\
=\frac{1}{\phi^2} \left[ 1- 4 \phi - \left( \frac{\partial}{\partial\phi}[\phi Z_N(\phi)] \right)^{-1} \right].
\end{eqnarray}
\noindent
With an analytic expression for the direct correlation function, we proceed to obtain the indirect correlation, $\gamma_{PYHD}(r)$, using the OZ relation Eq.~(\ref{eqOZ}) expressed in the Fourier space,
\begin{equation}
    h(q)=c(q) + \rho h(q) c(q) \; , \label{eqOZq}
\end{equation}
\noindent
where $h(q)$ and $c(q)$ are the 2D Fourier transforms of $h(r)$ and $c(r)$, respectively. 
From Eq.~(\ref{eqOZq})  and Eq.~(\ref{eqgamma}) follows
\begin{equation}
    \gamma(q) = \frac{\rho c(q)^2}{1 - \rho c(q)} \; .
\end{equation}
\noindent
Then, by back--transformation $\gamma_{PYHD}(r)$ is obtained, and the pair correlation function, $g_{PYHD}(r)$, is calculated using Eq.~(\ref{eqgamma}).
Here, it is important to remark that since $c_{PYHD}(x;\phi)$ is not the exact PY direct correlation function, the $g_{PYHD}(r)$ obtained does not satisfy $g_{PYHD}(r)=0$ for $r<2R$. This was fixed setting $g_{PYHD}(r)=0$ in this region.
The bridge function $B_{PYHD}$ is straightforwardly computed using Eq.~(\ref{BPY}).
The derivative with respect to $\phi$ of bridge function, needed for the VMHNC--PY scheme, is obtained using the expression
\begin{equation}
\frac{\partial B_{PY}(r)}{\partial \phi} = 
\frac{-\gamma(r)}{\gamma(r) + 1} \; \frac{\partial \gamma (r)}{\partial \phi} .
\end{equation}
\noindent
Here, the derivative of $\gamma (r)$ is obtained back--transforming  
\begin{equation}
    \frac{\partial \gamma (q)}{\partial \phi} = 
    \left( 
    -1 + \frac{1}{\left( 1 - \rho c(q)\right)^2}
    \right) \frac{\partial c(q)}{\partial \phi} \; ,
\end{equation}
\noindent
where the derivative of $c(q)$, in turn, is calculated by Fourier transforming $\partial c(r)/\partial \phi$.
Note that the derivative of $c(r)$ can be analytically performed, and contains a Dirac delta term, which should be
analytically Fourier transformed.

\subsection{LB: Parameterized ``Exact'' hard disks}
\label{subsec:LB}

For this reference system, we followed the procedure described in detail by Law and Buzza in the Appendix of Ref.~\onlinecite{law2009}.
They, following Guo and Riebel~\cite{guo2006}, start also from the Baus and Colot~\cite{baus1987} expression for $c(r)$, but use the accurate and simple expression proposed by Santos~\cite{santos1995} for the compressibility factor.
Then, by using the OZ relation, they obtain the pair correlation function, which is further corrected by generalizing to 2D the Verlet and Weis~\cite{verlet1972} scheme. Finally, they generalize also the Henderson and Grundke~\cite{henderson1975} proposal to obtain the cavity function $y(r)$. 
At this point, the bridge function could be calculated, according to Eq.~(\ref{eqB}), directly using $c(r)$, the corrected $g(r)$ and the parameterized $y(r)$. However, this would result in a discontinuity at $2R$ of $\gamma(r)$, and consequently of the bridge function.
For this reason, we used the OZ relation one more time, but now starting from the corrected $g(r)$, to obtain a continuous $\gamma_{HD}(r)$ and $B_{HD}(r)$.
The derivative with respect to $\phi$ of bridge function, needed for the VMHNC--LB scheme, is obtained using finite differences.
%%

%\bibliography{mr_bib} 
%\bibliographystyle{rsc} %aip


\end{document}